# EINFÜHRUNG: ANONYMITÄT. WISSEN GESTALTEN DURCH NICHTWISSEN

Paula Helm

Anonymität zu erzeugen bedeutet Verbindungen zu kappen. Ein häufiges Ziel in diesem Zusammenhang ist das Verhindern von Zurechenbarkeit.[1] Dieses Verhindern von Zurechenbarkeit kann problematisch sein, beispielsweise wenn es dazu führt, dass straffällige Personen unentdeckt bleiben. Allerdings kann Unzurechenbarkeit auch Schutz vor Diskriminierung spenden. In medizinischen, religiösen oder juristischen Belangen ist dies von elementarer Bedeutung. Wenn Personen Anonymität aktiv herstellen, so tun sie dies also zumeist, weil sie verhindern wollen, dass bestimmte, das heißt sensible und/oder kompromittierende Informationen über sie mit ihrer Identität in Verbindung gebracht werden. Dadurch, dass sie als Individuen in Bezug auf bestimmte Informationen über sie unerreichbar bleiben, können sie in Formen des Austausches eintreten, die für sie ansonsten unmöglich wären. Zu denken ist hier beispielsweise an Praktiken des Austausches in (selbstorganisierten) Therapiegruppen, an das Ausleben stigmatisierter sexueller Vorlieben, an die Rolle von Anonymität in den Performing Arts oder an politische Widerstandsbewegungen.

Angesichts der Mannigfaltigkeit von Beispielen, in denen personale Anonymität von Bedeutung ist, wundert es nicht, dass es vor allem diese personenbezogenen Dimensionen sind, welche in aktuellen Debatten über die zunehmende Prekarität von Anonymität angesichts neuer technischer Möglichkeiten der Vernetzung im Fokus stehen.[2] Hier wollen wir uns trotzdem – oder gerade deshalb – auf einen anderen Aspekt von Anonymität konzentrieren, der bisher noch sehr viel weniger Aufmerksamkeit erfahren hat. Und zwar gehen wir davon aus, dass das Forschen und Arbeiten mit und über Anonymität neue Perspektiven auf und für gegenwärtige Formen der Wissensproduktion eröffnen kann.[3] Damit schließen wir an einen Diskurs an, der vor

---

1 *Helen Nissenbaum:* The Meaning of Anonymity in an Information Age. In: The Information Society 15 (1999), S. 141–144.
2 *Paul Ohm:* Broken Promises of Privacy: Responding to the Surprising Failure of Anonymization, 57 UCLA Law Review (2010) Heft 57, S. 1701–1777; *Gabriella Coleman:* Epilgue. In: Hacker, Hoaxer, Spy. The Many Faces of Anonymous. New York 2014, S. 401–426.
3 Die Idee zu diesem Panel geht auf eine Kooperation zwischen zwei von der VW-Stiftung geförderten, interdisziplinären Forschungsprojekten zurück. Das waren zum einen das Projekt ›Reconfiguring Anonymity‹ und zum anderen das Projekt ›Strukturwandel des Privaten‹. Ursprünglich auf dem Panel vertreten waren unterschiedliche Beteiligte beider Projekte. Über die drei hier vertretenen AutorInnen Linda Monsees, Julien McHardy und Paula Helm hinaus waren dies noch Michi Knecht mit einem Beitrag zu Public Anthropology und künstlerisch-allem durch den britischen Wissenschaftstheoretiker und Techniksoziologen John Law angestoßen wurde.

In seinem ebenso einschlägigen wie provokanten Werk ›After Method‹ bringt Law ein methodologisches Unbehagen zum Ausdruck. Dieses Unbehagen ist damit zu begründen, dass in all unserem wissenschaftlichen Streben nach Rationalisierung, Klassifizierung, Entmystifizierung und Ordnung, Kostbares verloren zu gehen droht. Worauf Law hier verweist, ist all das, was sich nicht rationalisieren und ordnen lässt – sich diesen Tätigkeiten regelrecht entzieht.[4] Mit der Artikulation seines (wohlbegründeten) Unbehagens hat Law einen Diskurs angestoßen, in dessen Rahmen bis heute unterschiedliche Vertreter*innen der Science and Technology Studies, der Soziologie sowie der Kultur- und Sozialanthropologie das Anliegen voranbringen, aktuelle Limitationen akademischer Wissensproduktion zu überwinden.[5] Die programmatischen Leitfragen dieses Diskurses lauten wie folgt:

Welche neuen Zugänge und Konzepte benötigen wir, um auch Leidenschaftliches, Chaotisches, Flüchtiges, Irrationales und Willkürliches auf eine Weise wissenschaftlich zu thematisieren, die der ganz eigenen Bedeutung dieser Momente des Lebens gerecht wird? Wie können wir Formen der Wissensproduktion mobilisieren, die das ›Andere‹ der derzeit vornehmlich auf Klarheit und Präzision ausgerichteten Standards zu repräsentieren vermögen? Wie lässt sich dieses ›andere‹ Wissen für die Forschung fruchtbar machen?[6] Anonymität kann einen interessanten Beitrag im Zusammenhang mit der Diskussion dieser Fragen leisten, weil es bei ihr um die Herstellung von Situationen, Praktiken, Ästhetiken, Kollektiven und Infrastrukturen geht, die erst durch eine bestimmte Form des Nichtwissens und durch das Kappen gewohnter Verbindungslinien möglich werden. Das Nicht-Wissen ist hier also gegenüber dem Wissen nicht einfach nur ein Anderes, das es zu überwinden gilt, sondern es erhält eine aktive, produktive Rolle. Es ermöglicht nicht nur bestimmte Tätigkeiten, es eröffnet auch neue Perspektiven. Diese Produktivität von Nichtwissen stand im Zentrum des Interesses dieses Panels.

---

wissenschaftlicher Kollaboration im Forschen über Anonymität, Amelie Baumann mit einem Beitrag zu anonymer Samenspende und Götz Bachmann mit einem Kommentar.

4 *John Law:* After Method. Mess in Social Science Research. London/New York 2004.
5 Siehe hierzu beispielsweise die in *John Law/Evelyn Ruppert* (Hg.): Modes of Knowing. Resources from the Barock. Manchester 2016 versammelten Beiträge.
6 John Law bezieht sich dabei vor allem auf die Sozialwissenschaften, doch lässt sich seine Kritik problemlos auch auf viele Bereiche der Anthropologie/Ethnologie beziehen, siehe etwa *Roland Littlewood* (Hg.): Knowing and Not Knowing in the Anthropology of Medicine. London 2007 oder *Mattijs van de Port:* Baroque as Tension: Introducing Turmoil and Turbulence in the Academic Text. In: John Law/Evelyn Ruppert (Hg.): wie Anm. 5; sowie einige der Beiträge in *John Law/Annemarie Mol* (eds.): Complexities: Social Studies of Knowledge Practices. North Carolina 2002.

Die folgenden drei Beiträge greifen die Diskussion auf und beschäftigen sich aus unterschiedlichen Zugängen heraus mit einer etwaigen wissenschaftlichen Produktivität von Anonymität als Forschungsthema und Forschungsmodus. Dazu werden unterschiedliche Bereiche in den Blick genommen: Im ersten Beitrag untersucht Paula Helm einen Kontext, in dem mit Anonymität zum Zwecke alternativer Wissensproduktion experimentiert wird: den experimentellen Tanz. Im experimentellen Tanz wird Anonymität ganz dezidiert dafür eingesetzt, aus vorgefertigten Denk-, Bewertungs- und Bewegungsschemata auszubrechen. Anonymität soll hier dabei helfen, neue Möglichkeitsräume zu schaffen. Dabei geht es um Themen der Selbst- und Fremdwahrnehmung sowie um politische Projekte im Kontext postkolonialer, feministischer Bewegungen.

Linda Monsees diskutiert im zweiten Beitrag eine Teilnehmende Beobachtung, die sie im Rahmen sogenannter ›Cryptoparties‹[7] durchgeführt hat. Im Zentrum ihrer Auseinandersetzung steht dabei die Frage, mit welchen Mitteln hier ›anonymity-literacy‹, das heißt Anonymitätskompetenzen vermittelt werden, und welche (impliziten) politischen Positionen bei dieser nicht-hierarchischen Form der Wissensvermittlung in Kraft gesetzt werden.

Die beiden Stränge flechtet Julien McHardy in Hinblick auf die gemeinsame Fragestellung nach dem Verhältnis von Wissen, Nicht-Wissen und Anonymität in seinem Beitrag zusammen. Dabei konzentriert er sich auf die Trope des ›Sonnenuntergangs‹ und schlägt davon ausgehend vor, Anonymität als eine Intervention zu begreifen, mit welcher sich die Verhältnisse von Anwesenheit und Abwesenheit neu komponieren lassen.

Dr. Paula Helm
Interdisziplinäres Zentrum für Ethik in den Wissenschaften
Abteilung für Gesellschaft, Kultur und technischer Wandel (SCRATCH)
Wilhelmstr. 19
72074 Tübingen
paula.helm@uni-tuebingen.de

---

7 Das sind Graswurzelveranstaltungen, auf denen Ehrenamtliche interessierten Laien Anonymitätskompetenzen für die Nutzung neuer Informationstechnologien vermitteln.